\documentclass[aps,prl,twocolumn,showpacs,floatfix,superscriptaddress]{revtex4}
\usepackage{amsmath,amsbsy,amssymb,latexsym}
\usepackage{graphicx,float}
\usepackage{color}

\begin{document}
\title{Single temperature for Monte Carlo optimization on complex landscapes}

\author{Denis Tolkunov}
\affiliation{Department of Physics and Astronomy, Rutgers University, Piscataway, NJ 08854-8019}
\affiliation{BioMaPS Institute for Quantitative Biology, Rutgers University, Piscataway, NJ 08854-8019}

\author{Alexandre V. Morozov}
\thanks{Corresponding author: morozov@physics.rutgers.edu}
\affiliation{Department of Physics and Astronomy, Rutgers University, Piscataway, NJ 08854-8019}
\affiliation{BioMaPS Institute for Quantitative Biology, Rutgers University, Piscataway, NJ 08854-8019}

\date{\today}

\begin{abstract}
We propose a new strategy for Monte Carlo (MC) optimization on rugged multi-dimensional landscapes.
The strategy is based on querying the statistical properties of the landscape in order to find the temperature
at which the mean first passage time across the current region of the landscape is minimized.
Thus, in contrast to other algorithms such as simulated annealing (SA), we explicitly match the temperature schedule to the statistics
of landscape irregularities. In cases where this statistics is approximately the same over the entire landscape,
or where non-local moves couple distant parts of the landscape, single-temperature MC
will outperform any other MC algorithm with the same move set. We also find that in
strongly anisotropic Coulomb spin glass and traveling salesman problems, the only relevant statistics (which we use to assign a single MC temperature)
is that of irregularities in low-energy funnels.
Our results may explain why protein folding in nature is efficient at room temperatures.
\end{abstract}

\pacs{05.10.Ln,   
02.70.Tt,  
02.60.Pn,  
02.50.Ey  
}
\maketitle

Numerous problems in science and technology such as
protein structure prediction, evolution on fitness landscapes, stochastic dynamics of complex systems and machine learning
require efficient global optimization of multivariate objective functions or ``energies''. The objective function $U(\vec{\bf x})$
can be viewed as a multi-dimensional (multi-${\cal D}$) landscape in which a certain quantity (potential energy, free energy, cost function, likelihood of a model)
is assigned to every configuration $\vec{\bf x}$ of an arbitrary number ${\cal D}$ of discrete or continuous state variables.
The optimization task is then to find a global minimum (or maximum)
on arbitrary landscapes as efficiently as possible. Since exact global optimization methods are not available,
various empirical approaches have been devised. A popular class of algorithms is based on the Metropolis MC scheme \cite{Metropolis:1953}.
This class includes the SA algorithm \cite{Kirkpatrick:1983}, as well as simulated tempering \cite{Marinari:1992}, parallel
tempering \cite{Hansmann:1997}, replica exchange \cite{Hukushima:1996}, ensemble MC \cite{Hesselbo:1995,Dittes:1996} and multicanonical MC \cite{Berg:1991}.
Non-Metropolis schemes for global optimization such as random-cost \cite{Berg:1993}
and genetic algorithms \cite{Goldberg:1989} have also been developed.
Another class of algorithms enables more efficient exploration
of the novel regions of the configuration space by making adaptive changes to the landscape \cite{Barhen:1997,Wenzel:1999,Hamacher:2006,Cvijovic:1995}.

Unfortunately, the empirical nature of these algorithms makes it impossible to predict which approach would perform best on a given problem.
In addition, most algorithms depend on problem-dependent adjustable parameters such as the SA
cooling schedule. Here we address these concerns by proposing a universal guiding principle for analyzing global optimization problems.
Our interest is not only in developing efficient, parameter-free global optimization schemes,
but also in understanding how stochastic simulations run by nature
(such as protein folding at constant temperature driven by thermal fluctuations) appear to be so much simpler than corresponding human-designed
algorithms.

\begin{figure}[t]
  \includegraphics[width=3.4in]{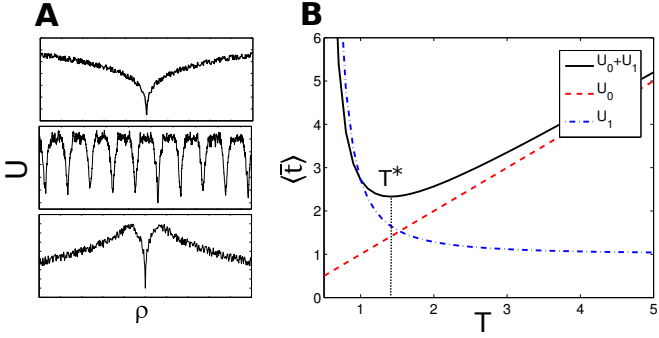}
  \caption{\label{Fig:theory} (Color online) A: Funnel, golf-course and misleading landscapes. B: $\langle \overline{t} \rangle$ as a
function of $T$ for $U_0 (x) = x$ and $U_1 = {\cal N} (0,1)$. Black (solid) curve: diffusion in $U_0 + U_1$, $\langle \overline{t} \rangle \sim T e^{1/T^2}$
and $T^\star = \sqrt{2}$. Red (dashed) curve: diffusion in $U_0$, $\overline{t} \sim T$ and $T^\star = 0$.
Blue (dashed-dotted) curve: diffusion in $U_1$, $\langle \overline{t} \rangle \sim e^{1/T^2}$ and $T^\star = \infty$.
}
\end{figure}

Our intuition is based on the notion of the \emph{global gradient} that leads towards good solutions (Fig.~\ref{Fig:theory}A, upper panel).
Landscapes without such a gradient are of the golf-course type or even the ``misleading'' type in which one has to go up before suddenly finding the
global minimum (Fig.~\ref{Fig:theory}A, middle and lower panels). In both of the latter scenarios it is necessary to sample ${\cal O} (N^{\cal D})$ possible states, where $N$
is the number of distinct values adopted by a (discretized) state variable.  
In contrast, global gradients define funnels on the landscape that can in principle be traversed in ${\cal O} (N)$ steps,
making efficient optimization possible.
A famous problem of this kind is protein folding \cite{Bryngelson:1995}, but any landscape in which gradual improvements
lead toward a good solution will have the funnel structure.
However, in realistic problems the global gradient will be weak and obscured
by the local ``noise'' or irregularities in the objective function (after all, in the absence of such noise any local optimizer would be successful).
As a result, the global gradient will be invisible at the smallest scale of a single MC step and can only be detected
from the average over a macroscopic local region. If this region is still relatively small, the global gradient will be approximately constant over it.
Furthermore, in strongly anisotropic problems the gradient may not be present everywhere but only in the low-energy valleys,
whereas the dominant high plateaus surrounding the valleys will be of the golf-course or misleading type.

Thus the global optimization problem can be formulated as diffusion (i.e., Metropolis MC sampling) in a potential which consists of random fluctuations with arbitrary
magnitude and correlation length superimposed onto a weak constant gradient.
Note that global optimization is different from computing thermodynamic properties,
which requires at least approximate equilibration and detailed balance. 
In contrast, global optimization is a strongly non-equilibrium process, with diffusion
at any point in the simulation affected only by the landscape features in its immediate neighborhood.

In the absence of random fluctuations, diffusion in a local region ${\bf L}$ subject to the constant force $\vec{\bf F} = -{\partial U}/{\partial \vec{\bf x}}$
is described by a multi-${\cal D}$ Fokker-Planck (FP) equation:
\begin{equation} \label{FP:multiD}
\frac{\partial \rho}{\partial t} = D \frac{\partial^2 \rho}{\partial \vec{\bf x}^2} - \vec{\bf v} \frac{\partial \rho}{\partial \vec{\bf x}},
\end{equation}
where $\rho(\vec{\bf x},t)$ is the probability distribution, $D$ is the diffusion coefficient, and $\vec{\bf v} = D \beta \vec{\bf F}$ is the
drift velocity ($\beta = 1/T$ is the inverse temperature).
We choose a coordinate system in which one of the axes ($x$) is parallel to $\vec{\bf F}$. In this system, Eq.~\eqref{FP:multiD}
factorizes into a one-dimensional ($1{\cal D}$) FP equation with a linear potential and ${\cal D}-1$ FP equations describing free diffusion.
We impose an absorbing boundary perpendicular to $\vec{\bf F}$ and focus on the $1{\cal D}$ FP equation with a linear potential.

The speed of propagation along $\vec{\bf F}$ is characterized by the mean first passage time (mfpt)
$\overline{t} (x,\beta)$, defined as the mean time required for a particle starting out at $x$ to reach the absorbing boundary $b$ for the first time.
With a reflecting boundary at $a$ ($a < x < b$), the mfpt is given by \cite{Lifson:1962,Weiss:1966}:
\begin{equation} \label{mfpt:1d}
\overline{t} (x,\beta) = \frac{1}{D} \int_x^{b} dx' e^{\beta U (x')} \int_a^{x'} dx'' e^{-\beta U (x'')},
\end{equation}
where $U(x)$ is the $1{\cal D}$ potential (objective function) in the direction of $\vec{\bf F}$.
For a linear potential in the absence of noise, $U_{0} (x) = F (b-x)~(F>0)$, we obtain:
\begin{eqnarray} \label{mfpt:1d:nonoise}
\overline{t}_{0} (x,\beta)\!=\!\frac{1}{D \beta F} \Big[\!(b -
x)\!+\! \frac{1}{\beta F} \left( e^{-\beta F (b - a)} - e^{-\beta
F (x - a)} \right)\!\Big]. \nonumber
\end{eqnarray}
In the $\beta \to 0$ (or $F \to 0$) limit the free-diffusion expression is recovered \cite{Lifson:1962}: $\overline{t}_{0} (x,0) = ({1}/{2D}) [(b-a)^2 + (x-a)^2]$.
However, for finite $\beta$ and $F$ we can always set $a$ so that the exponential terms on the right-hand side of Eq.~\eqref{mfpt:1d:nonoise} are
vanishingly small, eliminating the effect of the reflecting boundary on the diffusion process (formally, we take the $a \to -\infty$ limit):
\begin{equation} \label{mfpt:1d:nonoise:short}
\overline{t}_{0} (x,\beta) = \frac{b - x}{v},
\end{equation}
where $v = D \beta F$.
Now, assuming that the potential consists of the regular part and the irregular part, $U(x) = U_{0} (x) + U_{1} (x)$, and that
the characteristic length scale of $U_{1}$, $l_c$, is much smaller than the size ${\cal L}$ of the region over which $U_{0}$
is approximately linear, one can take a spatial average over the irregular part \cite{Zwanzig:1988}:
\begin{eqnarray} \label{mfpt:1d:noise:average}
\langle \overline{t} (x,\beta) \rangle &=& \frac{1}{D} \int_x^{b} dx' \int_{-\infty}^{x'} dx'' e^{\beta F (x'' - x')} \times \\
&& \int_{x' - L/2}^{x'+L/2} d x_1 e^{\beta U_1 (x_1)} \int_{x'' - L/2}^{x''+L/2} d x_2 e^{-\beta U_1 (x_2)}, \nonumber
\end{eqnarray}
where $l_c \ll L \ll {\cal L}$. Under these conditions and the additional assumption that $U_1$ statistics
does not change over ${\cal L}$, the two spatial averages are independent of each other and of $x', x''$, yielding
\begin{equation} \label{mfpt:1d:noise}
\langle \overline{t} (x,\beta) \rangle = H(\beta) \overline{t}_{0} (x,\beta),
\end{equation}
where 
$H(\beta) = \int_{-\infty}^{\infty} d U'_1 P(U'_1) e^{\beta U'_1} \int_{-\infty}^{\infty} d U''_1 P(U''_1) e^{-\beta U''_1}$
(we have switched from $x_1$ and $x_2$ to $U'_1 \equiv U_1(x_1)$ and $U''_1 \equiv U_1(x_2)$ in the spatial averages).
Furthermore, $H(\beta) = \int_{-\infty}^{\infty} d \Delta P(\Delta) e^{\beta \Delta}$, where
$P(\Delta)$ is the distribution of $\Delta = U'_1 - U''_1$ for $x_1, x_2$ constrained by $|x_1 - x_2| \gg l_c$. The last condition
guarantees that $P(\Delta)$ is independent of $|x_1 - x_2|$.

Clearly, if mfpt along $\vec{\bf F}$ is minimized for all local regions ${\bf L}$ with the constant gradient, the total time to
reach a good solution will also be minimized. The inverse temperature $\beta^\star$ that minimizes mfpt is given by:
\begin{equation} \label{mfpt:1d:min}
\left. \frac{d H(\beta)}{d \beta} \right\vert_{\beta^{\star}} = 
\frac{H(\beta^{\star})}{\beta^{\star}}.
\end{equation}
Note that $\beta^\star$ is independent of $F$.
Eq.~\eqref{mfpt:1d:min} can be used to find $\beta^\star$ numerically for any $P(\Delta)$.
If $P(U_1) = {\cal N} (0,\sigma^2)$, $P(\Delta) = {\cal N} (0,2\sigma^2)$
and $H(\beta) = e^{\beta^2 \sigma^2}$, yielding $\beta^{\star} = 1/\sqrt{2} \sigma$. With $T \ll T^{\star} = \sqrt{2} \sigma$
the diffusing particle gets stuck in local minima ($\langle  \overline{t} \rangle \sim e^{\sigma^2/T^2}$),
while for $T \gg T^{\star}$ diffusion is no longer optimally along the gradient of $U_0$ ($\langle  \overline{t} \rangle \sim T$) (Fig.~\ref{Fig:theory}B).
$\langle \overline{t} (x,\beta^{\star}) \rangle \sim \sigma/F$, indicating that diffusion is impeded by noise and aided by the gradient.
If $U_1 = 0$ everywhere, $P(\Delta) = \delta (\Delta)$ and $T^{\star} = 0$. Thus, as expected, the optimal solution in the absence of noise
is to roll down the potential at zero temperature. However, $\overline{t}_0 (x,\infty) = 0$ since Eq.~\eqref{mfpt:1d} does not accurately describe
the ballistic regime or strong forces. Finally, if $F \to 0$, $T^{\star} = \infty$ and $\langle  \overline{t} (x,0) \rangle$ reduces to the
expression for free diffusion, although any $T > \sqrt{2} \sigma$ will work almost as well (Fig.~\ref{Fig:theory}B).

Thus, if $U_1$ statistics is approximately constant and isotropic throughout the landscape, there is a unique MC temperature for
the most efficient minimization of the objective function (the anisotropic case will be presented elsewhere).
If not, different parts of the landscape are to be assigned different temperatures matched to the $U_1$ statistics.
All other schemes such as SA will yield suboptimal performance. In fact, if the amplitude of $U_1$
increases with decreasing $U_0$, our prescription calls for increasing the temperature as the simulation
progresses -- the exact opposite of the SA cooling schedule \cite{Kirkpatrick:1983}.

\begin{figure}[t]
  \includegraphics[width=2.2in]{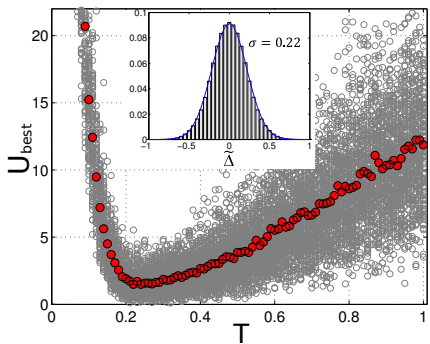}
  \caption{\label{Fig:TestFunctions} (Color online) Distribution of best predicted energies as a function of temperature for the 4D Griewank function
(Table~\ref{tab:TF}). For each $T$, $N_{\rm trials}$ independent trajectories with $N_{\rm iter}$ MC steps each 
were created by Metropolis MC sampling. The lowest energy $U_{\rm best}$ from each trajectory is shown as a grey dot.
Red dots are the average of $U_{\rm best}$ at a given $T$. Inset: Histogram of energy differences from a random sample of the landscape,
with a Gaussian fit (blue solid curve).
}
\end{figure}

Thus, in order to find the best MC temperature $T^{\star}$, we need to estimate $P(U_1)$ by sampling in the neighborhood of the current state
($P(U_1)$ can be resampled periodically during the simulation and
$T^{\star}$ recomputed). Unfortunately, it is difficult to extract $P(U_1)$ from $P(U)$ by detrending multi-${\cal D}$ walks.
Instead, we consider $P(\widetilde{\Delta})$, where $\widetilde{\Delta} = U(\vec{\bf x} + \Delta \vec{\bf x}) - U(\vec{\bf x})$ and $\Delta \vec{\bf x}$
is a single MC step with constant length. $|\Delta \vec{\bf x}|$ can be made so small
that the contribution from $U_0$ is negligible. With uncorrelated noise ($|\Delta \vec{\bf x}| > l_c$),
$P(\widetilde{\Delta}) = P(\Delta)$ and Eq.~\eqref{mfpt:1d:min} can be applied immediately.
However, if $|\Delta \vec{\bf x}| < l_c$, $U_1$ is smooth at the scale of a single MC step and
$P(\widetilde{\Delta})$ and $T^{\star}$ 
will depend on the move set. Indeed, $T^{\star} \sim |\Delta \vec{\bf x}|$
if the MC steps are so fine that $ U_1(\vec{\bf x} + \Delta \vec{\bf x}) - U_1(\vec{\bf x})$ is approximately linear. Nonetheless,
we find that for complex landscapes where $U(\vec{\bf x})$ is a sum of many independent terms,
$P(\widetilde{\Delta})$ quickly adopts a Gaussian shape if the sampling is over a region $\gg l_c$.
Since the MC walk is memoryless, $T^{\star} = \sqrt{2} \sigma$ still holds but now $\sigma$
depends on the move set. As $|\Delta \vec{\bf x}|$ increases beyond $l_c$, $\sigma$ converges to a universal value.

\begin{table}[h]
\begin{center}
\begin{tabular}{|c|c|c|c|c|c|}
\hline
\textbf{Function} & $N_{\rm iter}$ & $N_{\rm trials}$ & $T^{\star}_{\rm pred}$ & $T^{\star}_{\rm comp}$ & $U_{\rm best}^{\rm min} (T^{\star}_{\rm pred})$ \\
\hline \hline
\textbf{G} & 1.5$\times10^4$ & 100 & 0.22 & 0.22 & 0.010 \\
\hline
\textbf{R} & 5$\times10^3$ & 100 & 0.85 & 0.90 & 0.010 \\ 
\hline
\textbf{A} & 5$\times10^3$ & 100 & 0.40 & 0.45 & 0.008 \\
\hline
\end{tabular}
\caption{\label{tab:TF} Predicted and computed optimal temperatures for standard test functions \cite{Torn:1989}: the 4D Griewank (G) function
($U(\vec{\bf x}) = 1 + \frac{1}{4000} \sum_{i=1}^{4} x_i^2 - \prod_{i=1}^{4} \cos \big(\frac{x_i}{\sqrt{i}}\big)$, $x_i \in [-600,600], \forall i$),
the 4D Rastrigin (R) function ($U(\vec{\bf x}) = 4 + \sum_{i=1}^{4} \big(x_i^2 - \cos (18 x_i) \big)$, $x_i \in [-5,5], \forall i$) and
the 4D Ackley (A) function ($U(\vec{\bf x}) = 20 + e - 20 \exp \left( -0.2 \sqrt{\frac{1}{4} \sum_{i=1}^{4} x_i^2} \right) -
\exp \Big( \frac{1}{4} \sum_{i=1}^{4} \cos (2 \pi x_i) \Big)$, $x_i \in [-32.8,32.8], \forall i$). All three functions have multiple
local minima and a single global minimum located at $\vec{\bf x} = 0$ ($U(0) = 0$).
Each MC step is taken in a random direction and has a constant length of $1.0$ (G), $0.05$ (R) and $0.3$ (A).
$T^{\star}_{\rm pred}$ is based on a Gaussian fit to the histogram of $\widetilde{\Delta}$. $T^{\star}_{\rm comp}$ is the temperature at which
the average $U_{\rm best}$ is at minimum. $U_{\rm best}$ distribution at each $T$ was estimated using $N_{\rm trials}$
Metropolis MC runs of $N_{\rm iter}$ steps each. The lowest $U_{\rm best}$ at $T^{\star}_{\rm pred}$, $U_{\rm best}^{\rm min} (T^{\star}_{\rm pred})$, was
obtained using $N_{\rm iter} = 6\times10^4$ (G), $2\times10^4$ (R), and $10^4$ iterations (A),
for improved sampling of the tail of the $U_{\rm best} (T^{\star}_{\rm pred})$ distribution.
In all cases $\vec{\bf x}_{\rm best}^{\rm min}$ was within a single step from the global minimum.
}
\end{center}
\vspace{-0.6cm}
\end{table}

We have tested our approach on a set of standard functions often used to check performance of global optimization
algorithms \cite{Torn:1989} (Fig.~\ref{Fig:TestFunctions}, Table~\ref{tab:TF}). To estimate $P(\widetilde{\Delta})$, we
use $50$ trials with randomized starting positions and $10^5$ random steps each (all steps are accepted).
These parameters ensure that $P(\widetilde{\Delta})$ is close
to a Gaussian (with the rate of convergence dependent on the long-range order in the landscape
and on the complexity of the potential function),
and $T^{\star}$ is predicted as its $\sigma$. This simple procedure allows us to guess the best
temperature correctly (Table~\ref{tab:TF}), despite the fact that the landscapes are correlated and anisotropic
and the gradient is not guaranteed to be weak. Note that for Griewank and Rastrigin functions MC sampling yields a nearly flat region
around $T^{\star}_{\rm comp}$, making temperatures within a small range (e.g. $0.8$ to $1.0$ for the Rastrigin function) equally acceptable.

\begin{figure}[t]
  \includegraphics[width=3.4in]{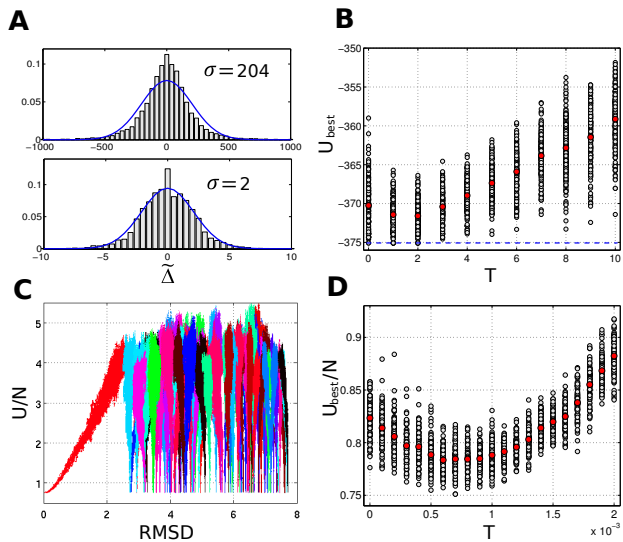}
  \caption{\label{Fig:Rugged} (Color online) A: $P(\widetilde{\Delta})$ estimated with unconstrained random walks ($10^2$ trials with
$10^4$ steps each) (upper panel) and with the funnel-sampling algorithm (lower panel).
B: Distribution of best predicted energies as a function of temperature for CSG ($N_{\rm trials} = 2\times10^2$, $N_{\rm iter} = 5\times10^5$).
The dashed horizontal line is the best energy found by extensive replica exchange runs \cite{Hukushima:1996,Hansmann:1997}.
This energy has been reached $2$, $6$ and $5$ times at $T = 0,1,2$ respectively.
C: Multi-funnel structure of the TS landscape.
$U/N$ is the average distance between neighboring cities in a given trajectory.
$100$ best minima were chosen from D, and for each minimum the funnel was mapped out
using $10$ random walks with $2\times10^4$ local steps each, and plotted in a distinct color.
Local steps involve exchanging two randomly picked neighboring cities.
The RMSD is computed with respect to the best solution in D.
D:  Distribution of best average distances between neighboring cities as a function of temperature for the TS problem ($N_{\rm trials} = 10^2$,
$N_{\rm iter} = 2.5\times10^5$).
}
\end{figure}

Next we turn to two more challenging global optimization problems, the Coulomb spin glas (CSG) \cite{Dittes:1996}
and the traveling salesmen (TS) problem \cite{Kirkpatrick:1983}. With CSG, we consider $N=50$ charges randomly distributed within the 3D unit cube:
$U(\vec{\bf s}) = \sum_{i=1, i \ne j}^{N} \sum_{j=1}^{N} {s_i s_j}/{|\vec{\bf{r}}_i - \vec{\bf{r}}_j|},$
where $s_i = \pm 1$ and the charge positions are fixed. A move involves flipping all signs in a randomly chosen subset of charges.

The CSG problem is characterized by the separation of scales: $P(\widetilde{\Delta})$ estimated using
unconstrained random walks (as was done for the test functions) yields a very high temperature,
since most of the landscape consists of high-energy plateaus that are either flat or have gradients pointing in random directions (Fig.~\ref{Fig:Rugged}A, upper panel).
MC runs at this temperature would not be able to utilize the global gradient information, which is restricted to low-energy funnels.
We therefore focus on the funnels to estimate $P(\widetilde{\Delta})$: 
from the current position with $U_{\rm cur}$, up to $N_{\rm m}$ ($2\times10^4$ for CSG) random moves are attempted.
If the new state is found with $U_{\rm new} \le U_{\rm cur}$, 
the loop terminates and the new state becomes the current state. Otherwise, the lowest energy among $N_{\rm m}$ new energies is chosen.
For CSG, we ran the algorithm $5$ times; each trajectory terminates once $10^3$ states have been accepted.
The resulting histogram (Fig.~\ref{Fig:Rugged}A, lower panel) correctly predicts the optimal temperature obtained by Metropolis MC
(Fig.~\ref{Fig:Rugged}B).
Its Gaussian shape suggests that $\widetilde{\Delta}$ distribution is isotropic in the funnels.
Surprisingly, even $T=0$ simulations yield reasonable results, indicating that
some deep funnels are smooth.

In the TS problem, one is given a list of cities and their locations, and the goal is to find the shortest possible tour that visits each city exactly once.
We considered $N=180$ cities randomly distributed within a $N^{1/2} \times N^{1/2}$ square, so that the average distance
between neighboring cities is independent of $N$ \cite{Kirkpatrick:1983}. We use Euclidean distances to compute $U$ and employ non-local moves in which
a segment of the trajectory is chosen at random and the direction in which all cities within that segment are traversed is inverted \cite{Kirkpatrick:1983}.
To reduce the degeneracy of low-scoring solutions, we start all trajectories from the same city. The TS landscape has a complex multi-funnel structure
(Fig.~\ref{Fig:Rugged}C) with high plateaus that dominate the landscape, so that only $\widetilde{\Delta}$ statistics within the funnels is relevant.
As in CSG, we employ the funnel-sampling algorithm (with $N_{\rm m} = 2\times10^3$)
to obtain $\sigma = 7.6\times10^{-4}$. This value is confirmed by scanning a range of temperatures with
fixed-temperature Metropolis MC (Fig.~\ref{Fig:Rugged}D).

Throughout this paper we have focused on minimizing mfpt. However, instead one may want to maximize the fraction of runs with $U_{\rm best}$
(the lowest energy from each MC trajectory)
below a certain cutoff. The tail of the $U_{\rm best}$ distribution at a given $T$ is affected by both its mean and standard deviation $\sigma'$, making
it possible that the temperature with the best mean is not the same as the temperature optimized for yielding extremely low-energy solutions.
However, from Fig.~\ref{Fig:TestFunctions} and Figs.~\ref{Fig:Rugged}B,D we see that $\sigma'$ varies with $T$ rather slowly. As a result, the
mfpt-based $T^{\star}$ remains valid, but in some cases the interplay between the mean and $\sigma'$ may make temperatures
in a small range around $T^{\star}$ equally acceptable.


If $U_1$ statistics is the same everywhere, global optimization is carried out
most efficiently by MC runs with a fixed temperature $T^{\star}$. However, if the nature of irregularities
changes across the landscape, two scenarios are possible. First, if $P(U_1)$ stays approximately the same
in regions $\gg \!{\bf L}$, the best temperature can be found for each region but needs to be updated as
the landscape is traversed. Mixing statistics from multiple regions will yield a single $T^{\star}$ that will not be the absolute best solution
but may still be a good approximate one.
Second, it is possible that different scales are mixed in a region $\ll \!{\bf L}$, e.g. due to anisotropy.
In this case $P (\widetilde{\Delta})$ will be non-Gaussian but Eq.~\eqref{mfpt:1d:min} still applies, yielding a single $T^{\star}$.
Using a single temperature works especially well with non-local steps such as those employed in the TS problem,
which can traverse a sizable part of the landscape in a single leap. Indeed, we find that even a multi-scale TS problem,
in which cities are clustered rather than randomly distributed  \cite{Kirkpatrick:1983}, has a unique best temperature
with non-local steps.

Our procedure can be viewed as an extension of the SA algorithm, which \textit{a priori} assumes that all scales are present in the problem
and, moreover, that they appear according to a specific cooling schedule. While SA may be the best way to proceed if the
properties of the landscape are completely unknown, quering some of the landscape statistics allows us to improve on the ``one size fits all''
SA technique by matching a given landscape to the appropriate temperature(s). We look forward to applying our
approach to protein structure prediction and other global optimization challenges.


This research was supported by National Institutes of Health (HG 004708) and by an Alfred P. Sloan Research Fellowship to AVM.


\end{document}